\documentstyle[amssymb,aps,preprint]{revtex}

\begin{document}
\title{Dynamics of Bright Soliton in Optical Fiber}
\author{Yi Tang and Wei Wang}
\address{National Laboratory of Solid State Microstructure and \\
Department of Physics, Nanjing University, Nanjing 210093, P. R. of China}
\date{\today}
\maketitle

\begin{abstract}
The bright soliton in optical fiber is generally investigated via its
spatial evolution in the time domain, where its waveform is considered in
many studies. To be consistent with the well-established picture of the
dynamics of solitons in other systems, in this letter, we propose it is
helpful to study the temporal evolution of the bright soliton by examining
its waveshape propagating along the space coordinate axis. We develop a
singular theory. Equations governing the evolution of the parameters of the
bright soliton in the slow time and the radiated field are explicitly
formulated for the first time. In addition, localized modes are found to
appear.
\end{abstract}


\hspace{1.2cm}

PACS (numbers): 03.40\vspace{0.3cm}.Kf, 52.35.Mw, 42.65.Tg


\widetext
%


Owing to the promising application to long distance soliton-based
communication and the great fundamental interest of physics of the process
involved, solitary waves and solitons in the nonlinear monomode optical
fiber have received intensive studies in recent years\cite{1,2}. The
generalized propagation equation of optical field in the fiber takes the form

\begin{equation}
iu_{x^{\prime }}^{\prime }+ik_1u_{t^{\prime }}^{\prime }-\frac 12%
k_2u_{t^{\prime }t^{\prime }}^{\prime }+\sigma \left| u^{\prime }\right|
^2u^{\prime }=i\varepsilon P^{\prime }\left[ u\right]  \eqnum{1}
\end{equation}
in which $x^{\prime }$ represents the propagation distance, $t^{\prime }$
the time and $u^{\prime }$ the complex field envelope. Usually, $\varepsilon
P^{\prime }\left[ u\right] $, including linear loss, high-order dispersion
and other nonlinear effects, is assumed to be small and treated as
perturbations to place emphasis on important phenomena of the bright and
dark solitons in the fiber\cite{3}. In the region of anomalous
group-velocity dispersion (GVD), by introducing the retarded time $T^{\prime
}=t^{\prime }-k_1x^{\prime }=t^{\prime }-x^{\prime }/v_g$, Eq. (1) is
normalized as

\begin{equation}
iu_{x^{\prime \prime }}^{\prime \prime }+\frac 12u_{TT}^{\prime \prime
}+\left| u^{\prime \prime }\right| ^2u^{\prime \prime }=i\varepsilon
P^{\prime \prime }\left[ u\right]  \eqnum{2}
\end{equation}
in terms of $T=T^{\prime }/T_0$, $x^{\prime \prime }=x^{\prime
}/L_D=x^{\prime }\left| k_2\right| /T_0^2$ and $u^{\prime \prime }=\sqrt{%
\left| k_2\right| /\sigma T_0^2}u^{\prime }$\cite{1}$.$ Customarily, Eq. (2)
is referred to as optical nonlinear Schr\"{o}dinger equation (NLSE), and its
unperturbed version supports distortionless propagation of a type of
solitary wave called the bright or temporal soliton\cite{3}.

Generally, waves travelling along the $x$-axis at speed $v$ are expressible
as functions of $(x-vt)$. A wave $F(x,t)$ may be thought of as formed from
the shape $f(\zeta )$ by the substitution $\zeta =(x-vt)$, or else as built
from the time signal $h(\tau )$ by the substitution $\tau =(t-x/v)$. Here,
the function $f(\zeta )$ with $f(x)=F(x,0)$ characterizes the ``waveshape'',
and $h(\tau )$ with $h(t)=F(0,t)$ depicts the ``waveform''\cite{4}.
Resulting pictures from the two standpoints for the wave $F(x,t)$ are that
the ``waveshape'' changes and propagates along the $x$-axis as time elapses
and the ``waveform'' distorts versus the retarded time $\tau $ as the
distance $x$ keep increasing. These actually presents two different point of
views for the visualization of scenario of soliton under perturbations.

The bright soliton propagating in the fiber governed by Eq. (2) was
typically investigated by interchanging the roles of the retarded time $T$
and the space $x^{\prime \prime }$ and defining an ``initial-value''
problem, or equivalently by directly treating the space $x^{\prime \prime }$
as the evolution coordinate and defining a boundary-value problem.
Accordingly, the aspect of waveform of the bright soliton was taken into
consideration and studies could benefit from the direct application\cite{5,6}
of the celebrated frameworks developed by Zakharov and Shabat (ZS) and
Ablowitz-Kaup-Newell-Segur (AKNS). Nevertheless, to avoid complication of
the ZS\ and AKNS schemes, other elaborate approaches were developed in the
framework of direct expansion as well\cite{7,8}.

In contrast to the studies of bright soliton, the aspect of waveshape is
extensively examined in other soliton problems with perturbations\cite{6},
including envelope soliton of the integrable cubic NLSE in water and other
applications\cite{9,10}. Although results for the understanding of the
waveform of bright soliton have been achieved, a natural question, how the
waveshape evolves in the real time or what the dynamics of bright soliton
is, is inevitable to arise. To answer this question, the corresponding
mathematical model is essentially different from the one investigated in
previous theories, and is also intractable in the ZS and AKNS schemes.
Consequently, a new theoretical challenge turns up. In this letter, we
introduce our theory for the subject.

Let's start from the dimensionless form of Eq. (1) in the anomalous
dispersion regime of the fiber

\begin{equation}
iu_x+iu_t+\frac 12u_{tt}+\left| u\right| ^2u=i\varepsilon P\left[ u\right] 
\eqnum{3}
\end{equation}
where $t=t^{\prime }/t_0$, $x=x^{\prime }/l$ $=x^{\prime }\left| k_2\right|
/t_0^2$, $u=\sqrt{\left| k_2\right| /\sigma t_0^2}u^{\prime }$ and $%
t_0=\left| k_2\right| /k_1=v_g^{-1}\left| dv_g/d\omega \right| .$ Obviously,
instead of the usual $T_0$ that is determined by the width of the input
waveform in existing theories, a characteristic time, namely $t_0$ that is
determined by the working wavelength and nature of the fiber, is used in the
normalization. Formally, Eq. (3) differs from Eq. (2) or the normal form of
optical NLSE only by an additional term $iu_t$ due to invalidation of the
retarded time, but the essential difference lies in that the time $t$ here
must be treated as the evolution coordinate and an initial-value problem is
consequently defined, since the waveshape of bright soliton is to be taken
into account. In the absence of perturbations, the bright soliton admitted
by Eq. (3) is given by

\begin{eqnarray}
u_{sol}(x,t) &=&2\eta {\rm sech}2\eta (2\zeta +1)[x-\frac 1{(2\zeta +1)}%
t-\chi ^{\prime }]  \nonumber \\
&&\times \exp \{-i[2(\zeta ^2-\eta ^2+\zeta )x-2\zeta t-\theta _1]\} 
\eqnum{4}
\end{eqnarray}
provided the initial waveshape is in the form 
\begin{eqnarray}
u_{sol}(x,0) &=&2\eta {\rm sech}2\eta (2\zeta +1)[x-\chi ^{\prime }] 
\nonumber \\
&&\times \exp \{-i[2(\zeta ^2-\eta ^2+\zeta )x-\theta _1]\}.  \eqnum{5}
\end{eqnarray}

From the general point of view of solitons under perturbations, if the
perturbations turn on, the bright soliton with a starting state of Eq. (5)
to propagate in the fiber governed by Eq. (3) can not be described by Eq.
(4), it undergoes a slow change in the time via the variation of its
parameters. Moreover, other wave modes come on to appear\cite{6,11}. To
characterize the picture, we introduce a slow time scale $t_1=\varepsilon t$
and assume that the solution of Eq. (3) is of the form

\begin{equation}
u(t,z,t_1)=[2\eta f(z)+\varepsilon v(t,z,t_1)]e^{-i\theta (z,t_1)}  \eqnum{6}
\end{equation}
where $f(z)={\rm sech}z$, $z=2\eta (2\zeta +1)(x-\varepsilon ^{-1}\chi -\chi
^{\prime })$ and $\theta =(Kz-\varepsilon ^{-1}\theta _0-\theta _1)$. Also,
we need further to assume that $\eta $, $\zeta $, $\chi $, $\chi ^{\prime }$%
, $K$, $\theta _0$, $\theta _1$ are dependent directly on $t_1.$ Obviously, $%
z$ is the coordinate variable in the reference frame tied up to the bright
soliton. Here, if we take $t$, $z$ and $t_1$ in place of $t$ and $x$ as new
independent variables, the derivatives with respect to time and space in Eq.
(3) are thus replaced by

\begin{eqnarray}
\frac \partial {\partial t}=\frac \partial {\partial t}-2 &&\eta (2\zeta
+1)\Lambda \frac \partial {\partial z}+\varepsilon \frac{\eta _{t_1}}\eta z%
\frac \partial {\partial z}+\varepsilon \frac{2\zeta _{t_1}}{(2\zeta +1)}z%
\frac \partial {\partial z}  \nonumber \\
&&-\varepsilon 2\eta (2\zeta +1)\chi _{t_1}^{\prime }\frac \partial {%
\partial z}+\varepsilon \frac \partial {\partial t_1}  \eqnum{7}
\end{eqnarray}
and

\begin{equation}
\frac \partial {\partial x}=2\eta (2\zeta +1)\frac \partial {\partial z} 
\eqnum{8}
\end{equation}
where $\chi _{t_1}=\Lambda $ is defined. Introducing Eqs. (6)-(8) into Eq.
(3), we transform from the laboratory frame into the soliton's one and get
two equations from $O(1)$ and $O(\varepsilon )$, respectively. Examining the
zeroth-order equation for $O(1)$, we derive $\Lambda =(2\zeta +1)^{-1}$, $%
\theta _{0t_1}=\Omega =2(\zeta ^2+\eta ^2)\Lambda $ and $K=(\zeta ^2-\eta
^2+\zeta )\Lambda \eta ^{-1}$. By virtue of these relations, we simplify the
first-order equation for $O(\varepsilon )$ as

\begin{equation}
\frac 12v_{tt}+i\Lambda ^{-1}v_t-2\eta v_{zt}+2\eta ^2v_{zz}+8\eta
^2h^2(z)v+4\eta ^2h^2(z)v^{*}-2\eta ^2v=R(z)  \eqnum{9}
\end{equation}
where asterisk $*$ denotes the complex conjugate$.$ And the ``source term''%
\cite{11} $R(z)=R_r(z)+iR_i(z)$ is given by

\begin{eqnarray}
R_r&=&-%
\mathop{\rm Im}%
(Pe^{i\theta })-4\eta (2\eta _{t_1}+\eta \Lambda \zeta _{t_1})\varphi _2(z) 
\nonumber \\
&+&2\eta \Lambda ^{-1}[2\eta \Lambda ^{-1}K\chi _{t_1}^{\prime }-4\eta
^2\chi _{t_1}^{\prime }+\theta _{1t_1}]\phi _1(z)  \nonumber \\
&-&2\eta [\Lambda ^{-1}(\eta ^{-1}\zeta _{t_1}-2\Lambda \eta _{t_1})-2(2\eta
\Lambda \zeta _{t_1}+\eta _{t_1})]\phi _2(z)  \nonumber \\
&+&16\eta ^3\Lambda ^{-1}\chi _{t_1}^{\prime }\phi _1^3(z)-8\eta [2\eta
\Lambda \zeta _{t_1}+\eta _{t_1}]z\phi _1^3(z)  \eqnum{10a}
\end{eqnarray}
and

\begin{eqnarray}
R_i &=&%
\mathop{\rm Re}%
(Pe^{i\theta })+4\eta ^2(\eta ^{-1}\zeta _{t_1}-\Lambda \eta _{t_1})\phi
_1(z)  \nonumber \\
&-&2[\Lambda ^{-1}\eta _{t_1}+4\eta ^2(\eta ^{-1}\zeta _{t_1}-\Lambda \eta
_{t_1})]\varphi _1(z)  \nonumber \\
&-&4\eta ^2[\theta _{1t_1}+2\eta \Lambda ^{-1}\chi _{t_1}^{\prime }K+\Lambda
^{-2}\chi _{t_1}^{\prime }]\varphi _2(z)  \eqnum{10b}
\end{eqnarray}
where $\phi _1(z)={\rm sech}z$, $\phi _2(z)=z{\rm sech}z$, $\varphi _1(z)=%
{\rm sech}z(1-z\tanh z)$, $\varphi _2(z)={\rm sech}z\tanh z$ are defined for
simplicity and later use. Expectably, a fresh equation comes out after the
linearization. Here, we should note that although the basic idea of the
present linearization is a natural extension of the normal scheme of
multiple scale expansion\cite{12}, the implementation of the idea in
handling such soliton problems as of the second order derivative with
respect to time is original. As usual, extra freedoms for the purpose of
preventing the occurrence of secular terms are introduced and included in
the source term. Taking advantage of Laplace transform to solve Eq. (9)
yields

\begin{equation}
\frac 12s^2\widetilde{v}+i\Lambda ^{-1}s\widetilde{v}-2\eta s\widetilde{v}%
_z+2\eta ^2\widetilde{v}_{zz}+8\eta ^2h^2(z)\widetilde{v}+4\eta ^2h^2(z)%
\widetilde{v}^{*}-2\eta ^2\widetilde{v}=s^{-1}R(z)  \eqnum{11}
\end{equation}
where $\widetilde{v}$ stands for the Laplace transform of $v$. Putting $%
v=v_1+iv_2$ and $\widetilde{v}=\widetilde{v}_1+i\widetilde{v}_2=(w_1+iw_2)e^{%
\frac{sz}{2\eta }}$, we derive from the real and imaginary parts of Eq. (11)

\begin{eqnarray}
sw_1+2\eta ^2\Lambda \widehat{L}_1w_2 &=&s^{-1}\Lambda R_ie^{-\frac{sz}{%
2\eta }}  \eqnum{12a} \\
sw_2-2\eta ^2\Lambda \widehat{L}_2w_1 &=&-s^{-1}\Lambda R_re^{-\frac{sz}{%
2\eta }}  \eqnum{12b}
\end{eqnarray}
where two Hermitian operators $\widehat{L}_1=$d$^2/$d$z^2+(2{\rm sech}^2z-1)$
and $\widehat{L}_2=$d$^2/$d$z^2+(6{\rm sech}^2z-1)$ are defined.

To solve Eq. (12) by virtue of eigen-expansion, a complete set of basis is
needed. Considering the homogeneous counterpart of Eq. (12), we derive the
following eigen-value problem

\begin{eqnarray}
\widehat{L}_1\phi &=&\lambda \varphi  \eqnum{13a} \\
\widehat{L}_2\varphi &=&\lambda \phi .  \eqnum{13b}
\end{eqnarray}
Now, if we define a non-Hermitian operator $\widehat{H}=\widehat{L}_2%
\widehat{L}_1$, then the corresponding adjoint operator is $\widehat{H}%
^{\dagger }=\widehat{L}_1\widehat{L}_2$. Using the operator $\widehat{L}_2$
to act on both sides of Eq. (13a) and then the $\widehat{L}_1$ on Eq. (13b)
gives

\begin{eqnarray}
\widehat{H}\phi &=&\lambda ^2\phi  \eqnum{14a} \\
\widehat{H}^{\dagger }\varphi &=&\lambda ^2\varphi .  \eqnum{14b}
\end{eqnarray}
Eigenstates of operators $\widehat{H}$ and $\widehat{H}^{\dagger }$ are
composed of a continuous spectrum with eigenvalue $\lambda =-(k^2+1)$ and
doubly degenerated discrete states with eigenvalue $\lambda =0,$
respectively. Under the definition of inner product in the Hilbert space,
Their eigenstates $\phi =\{\phi (z,k),\phi _1(z),\phi _2(z)\}$ and $\varphi
=\{\varphi (z,k),\varphi _1(z),\varphi _2(z)\}$ turn out to be a
biorthogonal basis (BB)\ with the completeness relation

\begin{equation}
\int_{-\infty }^{+\infty }\phi (z,k)\varphi ^{*}(z^{\prime },k)\text{d}%
k+\phi _1(z)\varphi _1(z^{\prime })+\phi _2(z)\varphi _2(z^{\prime })=\delta
(z-z^{\prime })  \eqnum{15}
\end{equation}
where

\begin{equation}
\phi (z,k)=\frac 1{\sqrt{2\pi }(k^2+1)}(1-2ik\tanh z-k^2)e^{ikz}  \eqnum{16}
\end{equation}
and

\begin{equation}
\varphi (z,k)=\frac 1{\sqrt{2\pi }(k^2+1)}(1-2{\rm sech}^2z-2ik\tanh
z-k^2)e^{ikz}  \eqnum{17}
\end{equation}
represent the continuous spectrum and $\phi _1(z)$, $\phi _2(z)$, $\varphi
_1(z)$, $\varphi _2(z)$ that are defined above stand for the discrete
states. BB is popular in the studies of non-Hermitian Hamiltonian problems%
\cite{13}. With the set of BB, we can expand the solutions of Eq. (12) as

\begin{eqnarray}
w_1(t,z,t_1) &=&\int_{-\infty }^{+\infty }\stackrel{\thicksim }{w}%
_1(t,k,t_1)\varphi (z,k)\text{d}k+\stackrel{\thicksim }{w}%
_{11}(t,t_1)\varphi _1(z)+\stackrel{\thicksim }{w}_{12}(t,t_1)\varphi _2(z) 
\eqnum{18a} \\
w_2(t,z,t_1) &=&\int_{-\infty }^{+\infty }\stackrel{\thicksim }{w}%
_2(t,k,t_1)\phi (z,k)\text{d}k+\stackrel{\thicksim }{w}_{21}(t,t_1)\phi
_1(z)+\stackrel{\thicksim }{w}_{22}(t,t_1)\phi _2(z).  \eqnum{18b}
\end{eqnarray}
Introducing Eq. (18) into Eq. (12) and solving by means of orthogonality of
the basis, we derive $w_1$ and $w_2$. Thus, $v_1$ and $v_2$ are determined
from the inverse Laplace transformation. Some terms directly proportional to 
$t$ and $t^2$ are found to appear in $v_1$ and $v_2$, they are non-physical
and called secular terms. But if we require

\begin{equation}
\int_{-\infty }^{+\infty }R_i(z)\phi _1(z)\text{d}z=0  \eqnum{19}
\end{equation}

\begin{equation}
\int_{-\infty }^{+\infty }R_i(z)\phi _2(z)\text{d}z+2\eta \Lambda
\int_{-\infty }^{+\infty }R_r(z)z\varphi _2(z)\text{d}z=0  \eqnum{20}
\end{equation}

\begin{equation}
\int_{-\infty }^{+\infty }R_r(z)\varphi _2(z)\text{d}z=0  \eqnum{21}
\end{equation}

\begin{equation}
\int_{-\infty }^{+\infty }R_r(z)\varphi _1(z)\text{d}z+2\eta \Lambda
\int_{-\infty }^{+\infty }R_i(z)z\phi _1(z)\text{d}z=0,  \eqnum{22}
\end{equation}
those terms vanish and we then get the final solution

\begin{eqnarray}
v_1 &=&\int_{-\infty }^{+\infty }\int_{-\infty }^{+\infty }\frac 1{2\eta
^2\lambda }(\sin \beta )R_i(z^{\prime })\phi ^{*}(z^{\prime },k)\varphi (z,k)%
\text{d}z^{\prime }\text{d}k  \nonumber \\
&&+\int_{-\infty }^{+\infty }\int_{-\infty }^{+\infty }\frac 1{2\eta
^2\lambda }(1-\cos \beta )R_r(z^{\prime })\varphi ^{*}(z^{\prime },k)\varphi
(z,k)\text{d}z^{\prime }\text{d}k  \nonumber \\
&&-\int_{-\infty }^{+\infty }\frac \Lambda {2\eta }R_i(z^{\prime })z^{\prime
}\phi _1(z^{\prime })\text{d}z^{\prime }\varphi _1(z)  \nonumber \\
&&-[\int_{-\infty }^{+\infty }\frac \Lambda {2\eta }R_i(z^{\prime
})z^{\prime }\phi _2(z^{\prime })\text{d}z^{\prime }+\int_{-\infty
}^{+\infty }\frac{\Lambda ^2}2R_r(z^{\prime })z^{\prime 2}\varphi
_2(z^{\prime })\text{d}z^{\prime }]\varphi _2(z)  \nonumber \\
&&+[\int_{-\infty }^{+\infty }\frac \Lambda {2\eta }R_i(z^{\prime })\phi
_2(z^{\prime })\text{d}z^{\prime }+\int_{-\infty }^{+\infty }\Lambda
^2R_r(z^{\prime })z^{\prime }\varphi _2(z^{\prime })\text{d}z^{\prime
}]z\varphi _2(z)  \eqnum{23}
\end{eqnarray}
and

\begin{eqnarray}
v_2 &=&-\int_{-\infty }^{+\infty }\int_{-\infty }^{+\infty }\frac 1{2\eta
^2\lambda }(\sin \beta )R_r(z^{\prime })\varphi ^{*}(z^{\prime },k)\phi (z,k)%
\text{d}z^{\prime }\text{d}k  \nonumber \\
&&+\int_{-\infty }^{+\infty }\int_{-\infty }^{+\infty }\frac 1{2\eta
^2\lambda }\{1-\cos \beta \}R_i(z^{\prime })\phi ^{*}(z^{\prime },k)\phi
(z,k)\text{d}z^{\prime }\text{d}k  \nonumber \\
&&+[\int_{-\infty }^{+\infty }\frac \Lambda {2\eta }R_r(z^{\prime
})z^{\prime }\varphi _1(z^{\prime })\text{d}z^{\prime }+\int_{-\infty
}^{+\infty }\frac{\Lambda ^2}2R_i(z^{\prime })z^{\prime 2}\phi _1(z^{\prime
})\text{d}z^{\prime }]\phi _1(z)  \nonumber \\
&&-[\int_{-\infty }^{+\infty }\frac \Lambda {2\eta }R_r(z^{\prime })\varphi
_1(z^{\prime })\text{d}z^{\prime }+\int_{-\infty }^{+\infty }\Lambda
^2R_i(z^{\prime })z^{\prime }\phi _1(z^{\prime })\text{d}z^{\prime }]z\phi
_1(z)  \nonumber \\
&&+\int_{-\infty }^{+\infty }\frac \Lambda {2\eta }R_r(z^{\prime })z^{\prime
}\varphi _2(z^{\prime })\text{d}z^{\prime }\phi _2(z)  \eqnum{24}
\end{eqnarray}
where we define $\beta =2\eta ^2\Lambda \lambda (t-\frac{(z^{\prime }-z)}{%
2\eta })$. It is noteworthy that localized modes turn out to appear in the
solution, which is essentially different from the envelope soliton of the
integrable cubic NLSE with the first-order temporal derivative\cite{14}.
From our viewpoint, the localized modes here is a kind of internal modes
whose occurrence is acknowledged to be intrinsic for nonintegrable models,
for instance, the $\phi ^4$ model\cite{15}. Although it is not sufficient to
conclude that the model we consider here is nonintegrable, the corresponding
Lax representation is really difficult to find. Returning to the restriction
condition imposed on the solution, we indicate that they can be satisfied by
the extra freedoms we introduce in advance. In fact, they result in a
sequence of novel equations

\begin{equation}
\eta _{t_1}=\frac \Lambda 2\int_{-\infty }^{+\infty }%
\mathop{\rm Re}%
(P\,e^{i\theta }){\rm sech}z\text{d}z  \eqnum{25}
\end{equation}

\begin{equation}
\zeta _{t_1}=-\frac \Lambda 2\int_{-\infty }^{+\infty }%
\mathop{\rm Im}%
(P\,e^{i\theta })\tanh z{\rm sech}z\text{d}z  \eqnum{26}
\end{equation}

\begin{eqnarray}
&&4\eta ^2[\Lambda ^{-2}+\frac 43\eta ^2]\chi _{t_1}^{\prime }  \nonumber \\
&=&\int_{-\infty }^{+\infty }%
\mathop{\rm Re}%
(Pe^{i\theta })z{\rm sech}z\text{d}z-2\eta \Lambda \int_{-\infty }^{+\infty }%
\mathop{\rm Im}%
(Pe^{i\theta })z\tanh z{\rm sech}z\text{d}z  \eqnum{27}
\end{eqnarray}

\begin{eqnarray}
&&2\eta [\Lambda ^{-1}-4\eta ^2\Lambda ]\times [\theta _{1t_1}+2\eta \Lambda
^{-1}K\chi _{t_1}^{\prime }]  \nonumber \\
&=&\int_{-\infty }^{+\infty }%
\mathop{\rm Im}%
(Pe^{i\theta }){\rm sech}z(1-z\tanh z)\text{d}z-2\eta \Lambda \int_{-\infty
}^{+\infty }%
\mathop{\rm Re}%
(Pe^{i\theta })z{\rm sech}z\text{d}z,  \eqnum{28}
\end{eqnarray}
which govern the dynamic evolution of bright soliton in the time. In
accordance with the usual definition of the width $w=1/2\eta (2\zeta +b)$,
we can derive a useful equation\newline
\begin{eqnarray}
w_{t_1} &=&4\eta ^2w^3\int_{-\infty }^{+\infty }%
\mathop{\rm Im}%
(P\,e^{i\theta })\tanh z{\rm sech}z\text{d}z  \nonumber \\
&&-w^2\int_{-\infty }^{+\infty }%
\mathop{\rm Re}%
(P\,e^{i\theta }){\rm sech}z\text{d}z.  \eqnum{29}
\end{eqnarray}

Now, we generate some specific results by examining two cases. At first, we
consider the linear loss given by $P\left[ u\right] =-\alpha _1u$. This
perturbation leads to $%
\mathop{\rm Re}%
(Pe^{i\theta })=-2\eta \alpha _1{\rm sech}z$. From Eq. (25), we compute $%
\eta _{t_1}=-2\alpha _1\eta \Lambda $, and then we obtain $\eta =\eta
_0e^{-2\alpha _1\Lambda t_1}=\eta _0e^{-2\varepsilon \alpha _1\Lambda t}$ by
integration. In this case, $\Lambda $ remains constant, thus, the
propagation distance of a fixed point of the soliton is calculated by $%
x=\Lambda t$. As a result, we can write

\[
\eta =\eta _0e^{-2\varepsilon \alpha _1x}, 
\]
which recovers a well-known result in previous theories\cite{1}.

Secondly, we give a brief study of the perturbation $P\left[ u\right]
=-i\alpha _2u\partial \left| u\right| ^2/\partial t$ accounting for the
Raman effect. Using Eq. (7), we derive $%
\mathop{\rm Im}%
(Pe^{i\theta })=-32\eta ^4\alpha _2\tanh z{\rm sech}^3z$, which has
influence on the soliton's width and velocity. By Eq. (29), we get $%
w_{t_1}=-8\alpha _2(2\eta )^6w^3/15$, integrating this equation yields

\[
w=w_0[1+\frac{16}{15}\alpha _2(2\eta )^6w_0^2t_1]^{-\frac 12}, 
\]
which exhibits that the soliton is narrowed under this effect. As well, we
can derive that the velocity decreases, obeying

\[
\Lambda =\Lambda _0[1+\frac{16}{15}\alpha _2(2\eta )^4\Lambda _0^2t_1]^{-%
\frac 12}. 
\]
Under the picture of waveform, the width is depicted differently, and the
velocity of dynamic sense can not be defined.

In conclusion, we think that the waveshape presents a more transparent
picture of directly physical significance than the waveform, especially in
the study of soliton under perturbations. Hence, we believe that our theory
is a nontrivial and necessary alternative for the subject. Moreover, the
mathematical development in this paper is distinct and normal, its idea is
helpful for the study of other soliton problems as well.

\vspace{0.3cm} {\sl {\bf Acknowledgments.}}This work was support by the NNSF
(No.19625409) and Nonlinear Project of the NSTC.





\end{document}